\documentclass[10pt,a4paper]{article}

\usepackage[latin1]{inputenc}
\usepackage{times}
\usepackage{graphicx}
\usepackage{amsmath}
\usepackage{amsfonts}
\usepackage[latin1]{inputenc}
\usepackage[T1]{fontenc}
\usepackage{verbatim}
\usepackage{latexsym}
\usepackage{natbib}

\usepackage{hyperref}

\hypersetup{backref,colorlinks=true,citecolor=blue}

\title{A method for filling gaps in solar irradiance \\ and in solar proxy data}

\author{Thierry Dudok de Wit \\ \normalsize LPC2E, Observatoire des Sciences de l'Univers en R\'egion Centre, \\ \normalsize 3A avenue de la Recherche Scientifique, 45071 Orl\'eans cedex 2, France \\
\normalsize (ddwit@cnrs-orleans.fr)}

\date{\normalsize \textit{Slightly expanded version of an article to appear in Astronomy and Astrophysics (2011), \\
Doi: 10.1051/0004-6361/201117024}}

\begin{document}
\maketitle

\begin{abstract}
Data gaps are ubiquitous in spectral irradiance data, and yet, little effort has been put into finding robust methods for filling them. We introduce a data-adaptive and nonparametric method that allows us to fill data gaps in multi-wavelength or in multichannel records. This method, which is based on the iterative singular value decomposition, uses the coherency between simultaneous measurements at different wavelengths (or between different proxies) to fill the missing data in a self-consistent way. The interpolation is improved by handling different time scales separately.

Two major assets of this method are its simplicity, with few tuneable parameters, and its robustness. Two examples of missing data are given: one from solar EUV observations, and one from solar proxy data. The method is also appropriate for building a composite out of partly overlapping records.
\end{abstract}

\section{Introduction}

Solar and stellar irradiance records are often plagued by data gaps. The proper interpolation of these missing data is a longstanding and notoriously delicate problem that requires a good understanding of the data \citep{wiener64,little02}. Considerable attention has been given to this problem in fields such as climate science \citep{dobesch07} but much less so in solar physics and in astrophysics. Often, the limited attention that is paid to data gaps contrasts with the sophistication of the analysis that is subsequently performed on these data. 

While short gaps can easily be filled by linear or by nonlinear interpolation, data gaps whose duration exceeds the characteristic time scales are much more difficult to handle. A notable exception is when multichannel synoptic observations of the same process are available, with gaps in some or in all of them. Spectral irradiance observations, which we shall concentrate on, precisely belong to that category. Our examples will be taken from the Sun, but the results can be easily extended to other types of multichannel observations. Our method applies to any set of observations that are recorded simultaneously (i.e. the time stamps are the same for all records), are correlated with each other, and whose time intervals fully or partly overlap. Our main assumption is their linear correlation, in the sense that each record can be approximated by a linear combination of the other ones.

A strong linear correlation is typically observed between spectral irradiance observations made at different wavelengths or between simultaneous measurements of different proxies. These synoptic records are frequently used to assess subtle changes in the variability of the Sun; they are often remarkably coherent in time $t$ and in wavelength $\lambda$. As a consequence, their variability can be explained in terms of a few contributions only. This property  is well known for the Extreme UltraViolet (EUV)  \citep{lean82,amblard08} but also for the visible range \citep{rabbette01}, when measured from space. The same coherency is observed among different proxies for solar activity \citep{pap92b,schmahl94,lean00,kane02c,floyd05,ddw09}. This property is rooted in the  structuring effect of the solar magnetic field. The coherency partly breaks down during the impulsive phase of solar flares because the spectrum then considerably depends on the local conditions of the solar atmosphere. Here, however, as in many applications, we consider daily or hourly averages, so that the effect of short transients can be discarded.

This coherency in both time and wavelength is the key to the reconstruction technique we shall introduce below. By interpolating along two dimensions (in time and along different records), we not only improve the quality of the reconstruction, but we also can fill arbitrarily large data gaps without having to rely on the tedious bookkeeping that is required by most interpolation schemes.

The nonparametric and data-adaptive method we advocate is based on the SVD or singular value decomposition \citep{golub00}, which is to linear algebra what the Fourier transform is to spectral analysis. The SVD allows the extraction of the coherent part of the solar spectral irradiance, which is then used to fill the data gaps iteratively. The method is described in Section 2, and two applications are detailed. The first one (Sec. 3) deals with solar spectral irradiance data in the EUV. In the second application (Sec. 4) we consider a set of solar proxies with numerous gaps.

\section{The reconstruction method}

Let $I(\lambda,t)$ be a multichannel record that represents either the solar spectral irradiance at different wavelengths (or in different spectral bands) or a set of solar proxies, or a combination thereof. All these quantities must be sampled simultaneously; the sampling rate, however, does not need to be constant. These data are conveniently stored in a matrix $\mathsf{I}_{ij} = [I(t_i,\lambda_j)]$, in which columns are time series. Each column may have an arbitrarily large number of data gaps, as long as a reasonable fraction of observations are available, say at least 20\%. 

\subsection{Basics}

The method we propose exploits either the coherency in wavelength, or both the coherency in wavelength and in time. We start with a description of the first option, because the second one can  be readily obtained by data embedding. Let us first assume that there are no gaps. The SVD of the data matrix then yields a separable set of functions (hereafter called \textit{modes})
\begin{equation}
I(t_i,\lambda_j) = \sum_{k=1}^M s_k \; u_k(t_i) \; v_k(\lambda_j) ,
\end{equation}
which are orthonormal 
\begin{equation}
\langle \langle u_k(t) u_l(t) \rangle = \langle v_k(\lambda) v_l(\lambda) \rangle =  \left\{
\begin{array}{lll}
0 & \textrm{if} & k \neq l \\
1 & \textrm{if} & k = l
\end{array} \right. .
\end{equation} 
The weights $s_1 \ge s_2 \ge \ldots \ge s_M \ge 0$ are positive by construction. The number $M$ of modes equals the rank of the matrix, which is usually the smallest of the number $N_t$ of samples or the number $N_{\lambda}$ of records. This decomposition is unique. The SVD of the data matrix directly yields a set of three matrices $\mathsf{I} = \mathsf{U} \mathsf{S} \mathsf{V}^T$ that respectively contain $u(t)$, the weights $s$, and $v(\lambda)$.

A key property of the SVD is that modes with heavy weights describe salient features of the data. That is, the truncated expansion
\begin{equation}
\hat{I}_K(t_i,\lambda_j) = \sum_{k=1}^{K \le M} s_k \; u_k(t_i) \; v_k(\lambda_j) 
\end{equation}
will capture the coherent part of the data while deferring incoherent fluctuations to the remaining modes. This property has made the SVD popular in multichannel and array data processing \citep{ddw95b,cline06}. We shall use it here to reconstruct the missing values.

The performance of the reconstruction can be quantified by the mean square error
\begin{equation}
e = \sum_{i=1}^{N_t} \sum_{j=1}^{N_{\lambda}} \left( I(t_i,\lambda_j) - \hat{I}_K(t_i,\lambda_j) \right)^2 = \sum_{k=K+1}^M s_k^2 ,
\end{equation}
which shows that by taking the few largest modes, the reconstruction error can be made arbitrarily small. As it turns out with spectral irradiance data, the first few weights are often orders of magnitude heavier than the subsequent ones, so that excellent reconstructions can be achieved with a few modes only. We implicitly assume here that features departing from the behaviour observed at other wavelengths are unlikely to have a solar origin (except during the impulsive phase of flares), so that they can be readily discarded. This will be illustrated below in Sec. 3.

Let us now assume that some samples are missing. The data covariance matrix and the SVD then cannot be computed anymore. This problem, however, can be circumvented by using the following iterative scheme with two embedded loops:
\begin{enumerate}
\item Fill each gap with some adequate value (typically the temporal mean of the record).
\item Compute the SVD.
\item Compute the approximation $\hat{I}_k$ of the data by retaining the $k$ largest mode(s) of the SVD. Initially, $k=1$.
\item Fill the gaps with $\hat{I}_k$, as defined in Eq. 3. As long as these values have not converged, go back to 2. (inner loop).
\item Increment the number of modes $k$ and start again at 2. Iterate until $k=K$ (outer loop).
\end{enumerate}

This method seems to have emerged independently in different contexts \citep{schneider01, beckers03, kondrashov06}; it has mostly been used for spatio-temporal data sets, with some subtle differences \citep{schneider07}. We refer to \citet{schneider01} for discussions on optimality, convergence, etc. Three additional adaptations, however, need to be considered before the method can be applied to irradiance data.

\subsection{Preprocessing problems}

The relative variability and the average value of the solar spectral irradiance vary by orders of magnitude between the soft X-ray and the visible range. The SVD, however, is scaling-dependent and so a renormalisation is required. We do so by standardising each record: first, the time average is subtracted and then a normalisation with respect to the standard deviation $\sigma_{\lambda_j}$ or the noise level (if known) is performed. Both operations are affected by the value of the missing samples, so they must be repeated at each iteration. This is particularly important for the offset subtraction. The renormalisation may be done only once.

\subsection{Multiscale decomposition}

The solar spectral variability contains a mix of scales that are driven by different processes: 27-day variations are due to solar rotation,  the 11-year periodicity is caused by the solar cycle, etc. Each of these processes leads to a specific spectral dependence; different scales should therefore be processed separately when filling gaps. This feature considerably improves the reconstruction skill and to the best of our knowledge has not yet been used.

The two ranges of scales that are most frequently encountered in solar studies are: below 81 days (which captures solar rotation and the evolution of active regions) and above 81 days. We apply the iterative SVD procedure described in Sec. 2.1 separately to both scales. The \textit{\`a trous} wavelet transform \citep{mallat08} is used to decompose the data into two records at each iteration: one with short time-scales and one with long time-scales. Classical bandpass filters may also be used because this has no significant impact on the results. The wavelet transform, however, is better suited for non-stationary data. One may also want to extract additional scales, such as the 13-day periodicity associated with centre-to-limb effects of hot coronal lines. This indeed results in a small but discernible improvement in the reconstruction of the EUV, at the expense of a longer computation time.

\subsection{Coherency in time}

The methodology so far only exploits the coherency between different wavelengths (or proxies), which is the key property. One may also want, however, to make use of the temporal coherency. This is useful when there are specific times at which there is no single observation, or if the number of records is small (typically $N_{\lambda}< 5$), or if each record can be considered as a smoothly varying waveform with incoherent noise superimposed on it.

The main asset of the iterative SVD reconstruction method is its straightforward extension to such a filtering in time, using the concept of embedding, which has been pioneered in the study of chaotic systems by \citet{broomhead86}. Let us expand the data matrix by appending replicates that are shifted in time, i.e.
\begin{equation}
\mathsf{E}_{ij} = \left[ I(t_i, \lambda_j) \ I(t_{i+1}, \lambda_j) \ I(t_{i+2}, \lambda_j) \ \ldots \ I(t_{i+D-1}, \lambda_j) \right] .
\end{equation}
By applying the SVD to this embedded matrix we exploit both the coherency in wavelength and in time. It is important (but not mandatory) that the data be regularly sampled since the method essentially computes a weighted average of each sample with its nearest neighbours. The higher the value of the embedding dimension $D$, the more adjacent time steps are used in the reconstruction, thus leading to a stronger smoothing in time. This is equivalent to using a symmetric finite-impulse filter whose coefficients are obtained data-adaptively. The particular case wherein one single record is embedded and decomposed by SVD is called singular spectrum analysis (SSA). In the SSA, only temporal information is used and so it is important for the embedding dimension to exceed the value of the dominant period in the data \citep{ghil02}. Our reconstruction, however, mostly relies on the strong coherency across wavelengths or proxies to fill the gaps and so the conditions on the value of the embedding dimension are much less stringent. In practice, low values ($D=2-5$) already bring a significant improvement.  The main reason for keeping this dimension as low as possible is to reduce the computational load.

\subsection{The method in practice}

The three tuneable parameters of the method are: a) the number $K$ of significant modes, b) the number of scales into which the data are decomposed, and c) the embedding dimension $D$. Only the first one really affects the outcome. A separation into two scales only (with a threshold between 50-100 days) is enough to properly capture both short- and long-term evolutions, and embedding dimensions of $D=2-5$ are usually adequate for reconstructing daily averages. The determination of the optimum parameters and the validation of the results is made by cross-validation and will be illustrated below.

The only critical question is memory and computational load. For an irradiance data set with five years of daily values at 100 wavelengths, and an embedding dimension of $D=5$, the size of the embedded matrix is $[1822, 500]$. The computation of the SVD at each iteration typically takes several seconds. For that reason, it may be desirable to process separately those spectral bands that evolve differently, such as the soft X-ray, the EUV and the MUV bands. The routine in Matlab$^{\mbox{\scriptsize{\textregistered}}}$  is available from the author.

\section{First example: gap filling in the EUV flux}

\begin{figure}
\centering
      \resizebox{0.65\hsize}{!}{\includegraphics{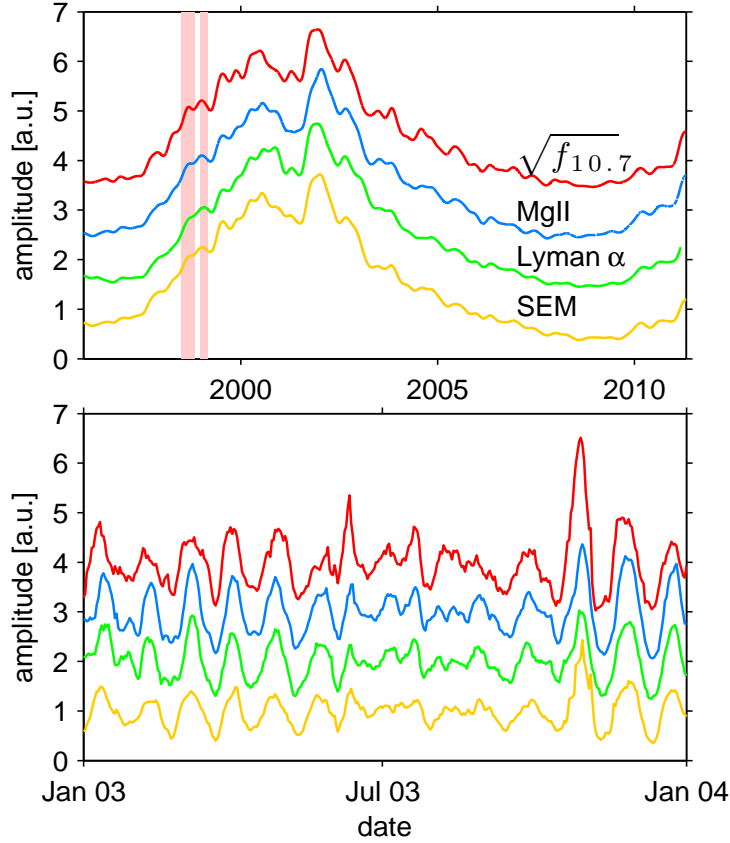}}
      \caption{Upper plot:  four chromospheric proxies, averaged over 80 days, using a Gaussian filter. The two major outages are shown shaded. Bottom plot: excerpt of the same proxies, showing daily values. The long-term trend has been subtracted from the latter. All records have been normalised to their standard deviation and shifted vertically for easier visualisation.}
         \label{fig_SEM_excerpt}
\end{figure}

The Solar EUV Monitor (SEM) is a solar Extreme UltraViolet (EUV) spectrometer that has been operating continuously on the SoHO satellite since January 1996 \citep{judge98}. In its first-order mode, SEM measures the irradiance within an 8 nm bandpass centred about the bright 30.38 nm He II line. On June 25, 1998, SoHO suffered a mission interruption, leading to the loss of several months of data. This long data gap considerably complicates the use of SEM data for upper atmosphere model validation. The SEM, however, mostly captures chromospheric emissions, which are highly correlated with other gauges of solar activity. Foremost among these are:
\begin{itemize}
\item the $f_{10.7}$ or decimetric index, which is the solar radio flux at 10.7 cm. This index, which is measured from the ground, captures a mix of thermal and electron gyro-resonance emissions, and has been shown to be highly correlated with the EUV flux \citep{tapping90}.
\item the Mg II index, which is the core-to-wing ratio of the Mg II line at 280 nm. This index is widely used as a proxy for chromospheric activity \citep{viereck01}. 
\item the intensity of the H I Lyman $\alpha$ line at 121.57 nm, which is the brightest spectral line below 200 nm \citep{woods00}.
\end{itemize}
Together with the flux from the SEM, we have four quantities that have different physical origins  and yet are highly correlated, thereby opening the prospect of filling the large gaps in the SEM data. We consider daily averages made from January 1, 1996 until April 29, 2011. The linear correlation between the $f_{10.7}$ index and the other proxies improves when taking its square root, which we shall systematically do from now on. The correlation between these four proxies on both long and short time-scales is illustrated in Fig.~\ref{fig_SEM_excerpt}.

\begin{figure}
\centering
      \resizebox{0.65\hsize}{!}{\includegraphics{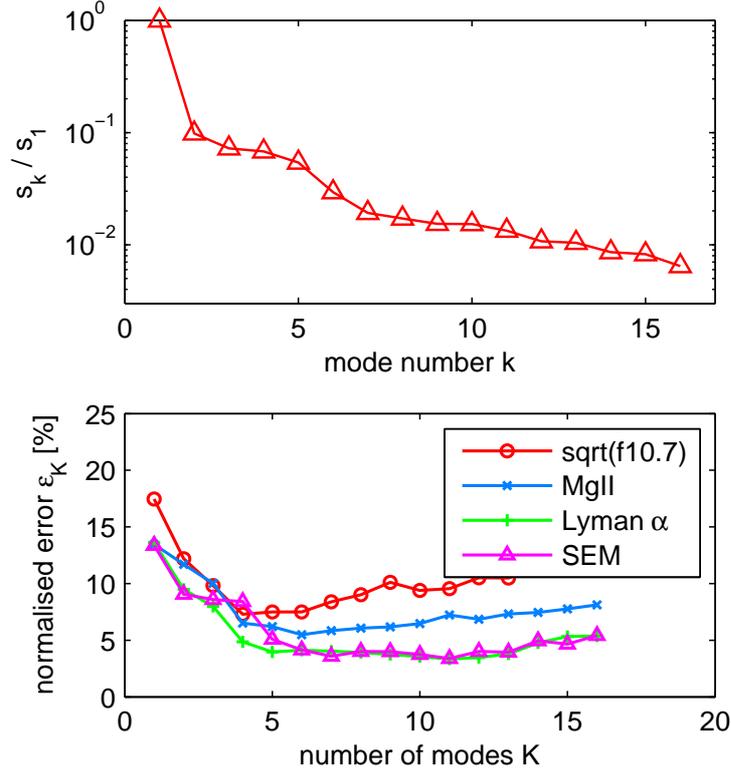}}
      \caption{Upper plot: distribution of the normalised weights $s_k/s_1$, for an embedding dimension of $D=4$. The total number of modes is 16. Bottom plot: variation of the reconstruction error with the number of modes $K$, for each of the four variables.}
         \label{fig_SEM_K}
\end{figure}

Our working hypothesis is that each of the missing samples from the SEM can be reconstructed from a linear combination of (possibly non-simultaneous) observations of the other proxies. As we shall see shortly, the best value of the embedding dimension is 4; let us therefore select $D=4$ and first determine the optimum number of modes. With four variables and an embedding dimension of 4, the total number of SVD modes is 16; their weights are displayed in Fig.~\ref{fig_SEM_K}. The first weight surpasses all the others because the first mode is an average of all four proxies, which is by far the most conspicuous coherent feature. The inflexion point between the few heaviest weights and the flat tail provides a convenient but visual criterion for determining the number of significant modes \citep{ddw95b}. According to this criterion, the best interpolation skill is  for $K=5-6$ modes out of 16. 

A better validation test consists in generating a small number of synthetic gaps, reconstructing them, and then checking how the residual error varies with the model parameters. To do so, we remove 5-10 \% of the samples from each record and then compute the normalised error
$$
\epsilon_K(\lambda_j) = \frac{1}{\sigma_{\lambda_j}} \sqrt{\frac{1}{N_{gaps}} \sum_{i=\{gaps\}} \left( I(t_i,\lambda_j) - \hat{I}_K(t_i,\lambda_j) \right)^2} ,
$$
where the average is computed for synthetic gaps only. This procedure is repeated ten times to obtain an estimate of the average value of the normalised error. A value of 100 \% can be interpreted as an error whose standard deviation equals the solar cycle variability of the original data. This value truly reflects the error made by filling short data gaps. Note that it tends to underestimate the error for larger gaps,  unless the length distribution of the synthetic gap  matches that of the original data.  

The evolution of the normalised error with $K$ is illustrated in Fig.~\ref{fig_SEM_K}, which shows a broad minimum around $K=4-8$, in agreement with the estimate obtained by visualisation. Note that the four minima occur at different values of $K$. The normalised error is on average larger for the $\sqrt{f_{10.7}}$ index, which suggests that this quantity is relatively more difficult to reconstruct than the others. This is not so surprising, because it is the only emission from the radio band. The smallest normalised error is obtained for the SEM, with $\epsilon_K = 4.5 \%$. This value is about half that of the estimated normalised uncertainty \citep{judge98}, which shows the excellent quality of the reconstruction. In practice, the optimum value of $K$ is frequently found to be one or two units higher than the value obtained by visual inspection. As Fig.~\ref{fig_SEM_K} suggests, an overestimation of $K$ is preferable to an underestimation.

The choice of the embedding dimension $D$ is mostly based on physical insight. With $D=1$ (i.e. no embedding) we assume that the missing samples are reconstructed from simultaneous observations only, whereas $D>1$ implies that the information contained in past and future observations is also used. Setting $D>1$ therefore involves a weighted averaging over time, which is appropriate for records whose samples are highly correlated in time. 

\begin{figure}
\centering
      \resizebox{0.65\hsize}{!}{\includegraphics{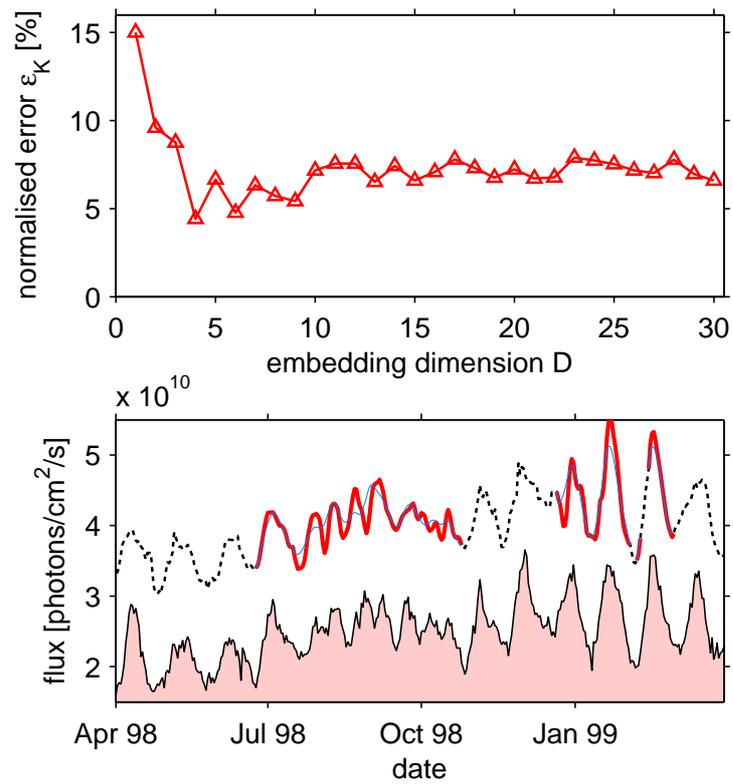}}
      \caption{Upper plot: variation of the reconstruction error for the SEM with the embedding dimension $D$. For each embedding dimension, the number $K$ of modes that minimises the error is chosen. Bottom plot: comparison between the measured flux from the SEM (dashed line) and the flux reconstructed with an embedding dimension of $D=4$ (thick line), and $D=15$ (thin line). The Mg II index is shown for comparison (filled curve), with arbitrary units.}
         \label{fig_SEM_D}
\end{figure}

In Fig.~\ref{fig_SEM_D} we estimate the normalised error for different embedding dimensions, using the optimum number of modes for each of them. The smallest error is obtained for an embedding dimension of $D=4$. Larger dimensions hardly reduce the error but do increase the computational load substantially. As expected, the higher the value of $D$, the smoother the reconstruction and the more likely that fine features may be missed. This is particularly evident in August 1998, when a group of rapidly evolving active regions were moving across the solar disc. An embedding dimension of 4 properly captures their evolution, whereas a dimension of 15 smears out all but the most pronounced peaks.

This example illustrates a relatively simple case because only one record has gaps in it. Let us now, however, consider a more frequent case in which several of the records have large gaps. Filling these gaps by standard interpolation schemes can become very time-consuming because of the amount of bookkeeping that is required to test whether gaps occur simultaneously in several records, etc. The SVD-based interpolation does not require any of these tests.

\section{Second example: reconstruction of the Ca K index}

The Ca K index is the normalised intensity of the Ca II K-line at 393.37 nm and has been advocated as a proxy for magnetic activity, including plages, faculae, and the network. This line is measured from the ground, so it cannot be observed continuously. Here we consider a record of daily observations made at the National Solar Observatory at Sacramento Peak \citep{keil98}, in which about  66\% of the samples are missing. This index is known to be highly correlated with other solar indices,  in particular with the Mg II index \citep{foukal09}, so that the SVD method is ideally suited for filling its gaps.

To reconstruct the missing values, we consider the following set of proxies that are highly correlated with the Ca K index: the square root of the $f_{10.7}$ index, the intensity of the H I Lyman ${\alpha}$ line, the Mg II index and the magnetic plage strength index (MPSI) \citep{parker98}. The time interval ranges from Nov. 1, 1980 to July 1, 2010; all proxies  have data gaps except for the first two. These gaps occur erratically and 6\% of them exceed 10 days. In this particular example, the coherency between proxies is crucial and indeed the choice of the embedding dimension $D$ does not significantly affect the results. Let us take $D=2$, which is the value that is recommended by the reconstruction error. The maximum number of SVD modes is 10 because we have five records. Out of these, three only are found to be significant.

\begin{figure}
\centering
      \resizebox{0.65\hsize}{!}{\includegraphics{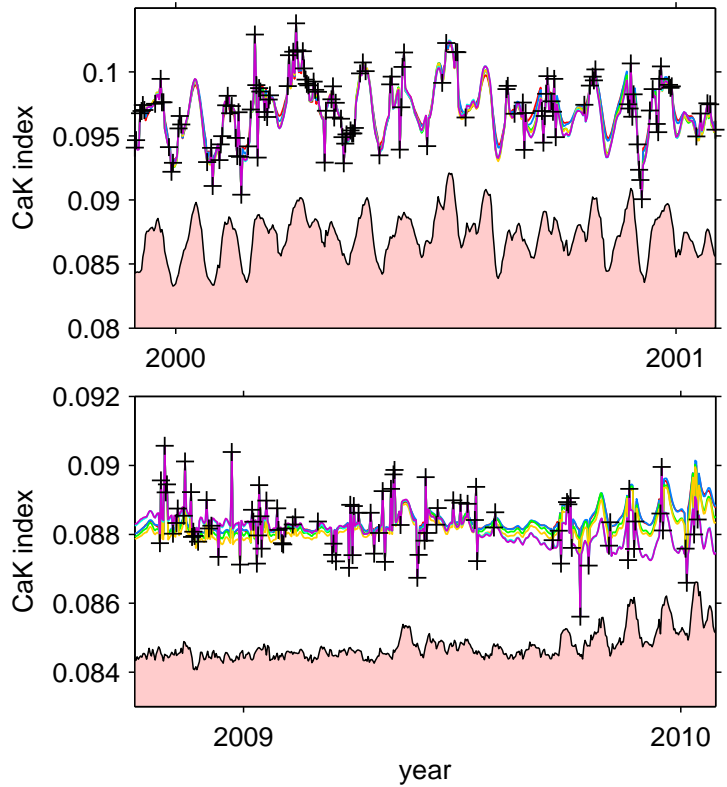}}
      \caption{Reconstruction of the missing values of the Ca K index using 1 to 6 modes. The upper plot shows an excerpt at solar maximum and the bottom one at solar minimum. The observations are indicated with crosses and the different reconstructions with continuous lines. Also shown is the Mg II index (filled line), in arbitrary units.}
         \label{fig_CaK1}
\end{figure}

The result of the reconstruction is illustrated in Fig.~\ref{fig_CaK1} for periods of high and low solar activity. Note that the results obtained with different number of modes lead to similar temporal evolutions. The reconstruction at solar maximum looks reasonable because it passes through the observations while staying highly correlated with the Mg II index. During solar minimum, however, the observed values of the Ca K index continue to fluctuate whereas the reconstructed values and the other proxies stay almost constant. The difference between the observed Ca K index and the smoothly varying reconstruction varies randomly in time, which questions its solar origin.

\begin{figure}
\centering
   \resizebox{0.65\hsize}{!}{\includegraphics{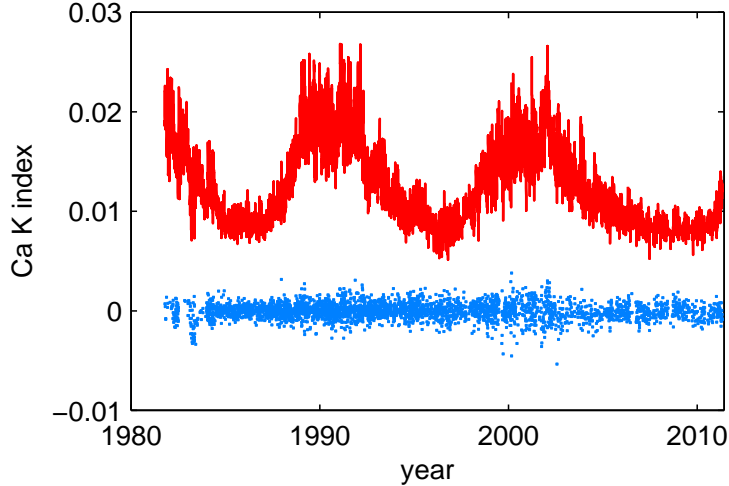}}
      \caption{Ca K index after reconstruction and filtering by wavelet transform (continuous curve) and the difference with the observations (crosses). For easier visualisation, the Ca K index has been shifted downwards by 0.08.}
         \label{fig_CaK2}
\end{figure}

To further investigate the origin of this difference between the observed and reconstructed index, we filtered the reconstructed data with the \textit{\`a trous} wavelet transform, which allows the separation of the sharp peaks from the more regular reconstruction. The residuals, i.e. the difference between the filtered reconstruction and the original observations, are shown in Fig.~\ref{fig_CaK2}: they are found to be independent and their Gaussian distribution only weakly varies with the solar cycle. This is a strong indication that the residuals are measurement errors rather than solar fluctuations. Their standard deviation is 0.0008, which represents 20 \% of the solar cycle variability of the Ca K index. Our reconstruction thereby provides a means for fitting the numerous data gaps in the Ca K index while also evaluating the confidence interval of the observations.

\section{Conclusions and additional applications}

This study shows that SVD-interpolation is a powerful technique for filling arbitrarily large gaps in multi-wavelength, multichannel or in synoptic records. We focused here on solar spectral irradiance observations, which are frequently plagued by missing data.  These gaps may be distributed at random in time or in wavelength. The main tuneable parameter is the number of SVD modes that is needed to reconstruct the data; this value may be estimated either by visualisation or by cross-validation. The method works best when each record can be approximated by a linear combination of the others. Since it relies on linear combinations only, it may be desirable to apply  a nonlinear static transform beforehand to increase the linear correlation between the records. 

For the method to work, the observations must be sampled simultaneously but not necessarily evenly. Non-simultaneous observations can be handled by resampling all variables to a common grid, for example by Fourier decomposition \citep[e.g.][]{hocke09}, and then filling the gaps by SVD. By alternating between the two, both the gaps and the interpolated values can be progressively refined.

This method has several applications in addition to mere interpolation. The first one is the cross-calibration of measurements of the same quantity by different instruments. The Mg II index, for example, is at present measured by different instruments that give different amplitudes. These  data sets are incomplete and only partly overlap, which considerably impairs their inter-comparison. The iterative SVD method is ideally suited for filling these gaps because the records are by definition strongly correlated.

A second potential application is the stitching together of total solar irradiance (TSI) observations. Merging TSI records from several instruments is a delicate and controversial task \citep{froehlich02} because instruments disagree on the absolute value of the TSI and often do not operate simultaneously. The iterative SVD provides a means for estimating the different offsets in a self-consistent way because it allows us to extrapolate each TSI record by assuming that its statistical properties with respect to the other records do not change in time. This property is particularly useful for checking composites that are built from different records, such as the TSI, the H I Lyman $\alpha$ intensity, the Mg II index and the sunspot index \citep{clette07}. This will be detailed in a forthcoming publication.


\subsubsection*{Acknowledgements}
{\small I thank the following institutes for providing the data: the Laboratory for Atmospheric and Space Physics (University of Colorado) for the Mg II and H I Lyman $\alpha$ composites, the National Solar Observatory at Sacramento Peak (data produced cooperatively by NSF/NOAO, NASA/GSFC and NOAA/SEC) for the Ca K index, the Mount Wilson Observatory (operated by UCLA, with funding from NASA, ONR  and NSF, under agreement with the Mt. Wilson Institute) for the MPSI index and the Space Sciences Center (University of Southern California) for the SEM data.  This study received funding from the European Community's Seventh Framework Programme (FP7/2007-2013) under the grant agreement nr. 218816 (SOTERIA project, http://www.soteria-space.eu).}


\begin{thebibliography}{}

\bibitem[{Amblard} et~al., 2008]{amblard08}
{Amblard}, P., {Moussaoui}, S., {Dudok de Wit}, T., {Aboudarham}, J.,
  {Kretzschmar}, M., {Lilensten}, J., and {Auch{\`e}re}, F. (2008).
\newblock {The EUV Sun as the superposition of elementary Suns}.
\newblock {\em Astron. Astroph.}, 487:L13--L16.

\bibitem[{Beckers} and {Rixen}, 2003]{beckers03}
{Beckers}, J.~M. and {Rixen}, M. (2003).
\newblock {EOF Calculations and Data Filling from Incomplete Oceanographic
  Datasets*}.
\newblock {\em Journal of Atmospheric and Oceanic Technology}, 20:1839--1856.

\bibitem[{Broomhead} and {King}, 1986]{broomhead86}
{Broomhead}, D.~S. and {King}, G.~P. (1986).
\newblock {Extracting qualitative dynamics from experimental data}.
\newblock {\em Physica D Nonlinear Phenomena}, 20:217--236.

\bibitem[{Clette} et~al., 2007]{clette07}
{Clette}, F., {Berghmans}, D., {Vanlommel}, P., {van der Linden}, R.~A.~M.,
  {Koeckelenbergh}, A., and {Wauters}, L. (2007).
\newblock {From the Wolf number to the International Sunspot Index: 25 years of
  SIDC}.
\newblock {\em Adv. Space Research}, 40:919--928.

\bibitem[Cline and Dhillon, 2006]{cline06}
Cline, A.~K. and Dhillon, I.~S. (2006).
\newblock {Computation of the singular value decomposition}.
\newblock In Hogben, L., editor, {\em Handbook of linear algebra}, chapter~45,
  pages 1--13. CRC Press, Boca Raton.

\bibitem[{Dobesch} et~al., 2007]{dobesch07}
{Dobesch}, H., {Dumolard}, P., and {Dyras}, I., editors (2007).
\newblock {\em Spatial Interpolation for Climate Data: The Use of GIS in
  Climatology and Meteorology}.
\newblock Geographical Information System Series. Wiley, London.

\bibitem[{Dudok de Wit}, 1995]{ddw95b}
{Dudok de Wit}, T. (1995).
\newblock {Enhancement of multichannel data in plasma physics by biorthogonal
  decomposition}.
\newblock {\em Plasma Physics and Controlled Fusion}, 37:117--135.

\bibitem[{Dudok de Wit} et~al., 2009]{ddw09}
{Dudok de Wit}, T., {Kretzschmar}, M., {Lilensten}, J., and {Woods}, T. (2009).
\newblock {Finding the best proxies for the solar UV irradiance}.
\newblock {\em Geoph. Res. Lett.}, 36:10107.

\bibitem[{Floyd} et~al., 2005]{floyd05}
{Floyd}, L., {Newmark}, J., {Cook}, J., {Herring}, L., and {McMullin}, D.
  (2005).
\newblock {Solar EUV and UV spectral irradiances and solar indices}.
\newblock {\em Journal of Atmospheric and Solar-Terrestrial Physics}, 67:3--15.

\bibitem[{Foukal} et~al., 2009]{foukal09}
{Foukal}, P., {Bertello}, L., {Livingston}, W.~C., {Pevtsov}, A.~A., {Singh},
  J., {Tlatov}, A.~G., and {Ulrich}, R.~K. (2009).
\newblock {A century of solar CaII measurements and their implication for solar
  UV driving of climate}.
\newblock {\em Solar Physics}, 255:229--238.

\bibitem[{Fr\"ohlich}, 2002]{froehlich02}
{Fr\"ohlich}, C. (2002).
\newblock The total solar irradiance variations since 1978.
\newblock {\em Adv. Space Research}, 29:1409--1416.

\bibitem[{Ghil} et~al., 2002]{ghil02}
{Ghil}, M., {Allen}, M.~R., {Dettinger}, M.~D., {Ide}, K., {Kondrashov}, D.,
  {Mann}, M.~E., {Robertson}, A.~W., {Saunders}, A., {Tian}, Y., {Varadi}, F.,
  and {Yiou}, P. (2002).
\newblock {Advanced Spectral Methods for Climatic Time Series}.
\newblock {\em Reviews of Geophysics}, 40:1003.

\bibitem[Golub and {Van Loan}, 2000]{golub00}
Golub, G.~H. and {Van Loan}, C.~F. (2000).
\newblock {\em Matrix Computations}.
\newblock Johns Hopkins Press, Baltimore.

\bibitem[{Hocke} and {K{\"a}mpfer}, 2009]{hocke09}
{Hocke}, K. and {K{\"a}mpfer}, N. (2009).
\newblock {Gap filling and noise reduction of unevenly sampled data by means of
  the Lomb-Scargle periodogram}.
\newblock {\em Atmospheric Chemistry \& Physics}, 9:4197--4206.

\bibitem[{Judge} et~al., 1998]{judge98}
{Judge}, D.~L., {McMullin}, D.~R., {Ogawa}, H.~S., {Hovestadt}, D., {Klecker},
  B., {Hilchenbach}, M., {Mobius}, E., {Canfield}, L.~R., {Vest}, R.~E.,
  {Watts}, R., {Tarrio}, C., {Kuehne}, M., and {Wurz}, P. (1998).
\newblock {First solar EUV irradiances obtained from SOHO by the CELIAS/SEM}.
\newblock {\em Solar Physics}, 177:161--173.

\bibitem[{Kane}, 2002]{kane02c}
{Kane}, R.~P. (2002).
\newblock {Correlation of solar indices with solar EUV fluxes}.
\newblock {\em Solar Physics}, 207:17--40.

\bibitem[{Keil} et~al., 1998]{keil98}
{Keil}, S.~L., {Henry}, T.~W., and {Fleck}, B. (1998).
\newblock {NSO/AFRL/Sac Peak K-line Monitoring Program}.
\newblock In {Balasubramaniam}, K.~S., {Harvey}, J., and {Rabin}, D., editors,
  {\em Synoptic Solar Physics}, volume 140 of {\em Astronomical Society of the
  Pacific Conference Series}, pages 301--308.

\bibitem[{Kondrashov} and {Ghil}, 2006]{kondrashov06}
{Kondrashov}, D. and {Ghil}, M. (2006).
\newblock {Spatio-temporal filling of missing points in geophysical data sets}.
\newblock {\em Nonlinear Processes in Geophysics}, 13:151--159.

\bibitem[{Lean}, 2000]{lean00}
{Lean}, J.~L. (2000).
\newblock {Short Term, Direct Indices of Solar Variability}.
\newblock {\em Space Science Reviews}, 94:39--51.

\bibitem[{Lean} et~al., 1982]{lean82}
{Lean}, J.~L., {Livingston}, W.~C., {Heath}, D.~F., {Donnelly}, R.~F.,
  {Skumanich}, A., and {White}, O.~R. (1982).
\newblock {A three-component model of the variability of the solar ultraviolet
  flux 145-200 nm}.
\newblock {\em Journal of Geophysical Research}, 87:10307--10317.

\bibitem[{Little} and {Rubin}, 2002]{little02}
{Little}, R.~J.~A. and {Rubin}, D.~B. (2002).
\newblock {\em {Statistical analysis with missing data}}.
\newblock Wiley series in probability and statistics. Wiley, New York, 2nd
  edition.

\bibitem[Mallat, 2008]{mallat08}
Mallat, S. (2008).
\newblock {\em {A Wavelet Tour of Signal Processing: the Sparse Way}}.
\newblock Academic Press, London, 3rd edition.

\bibitem[{Pap} and {Guhathakurta}, 1992]{pap92b}
{Pap}, J. and {Guhathakurta}, M. (1992).
\newblock {Variability of Solar UV Irradiance and Its Relation to the
  Variability in Coronal Green Line Index and Equivalent Width of He Line at
  1083nm}.
\newblock In {K.~L.~Harvey}, editor, {\em The Solar Cycle}, volume~27 of {\em
  Astronomical Society of the Pacific Conference Series}, pages 483--490.

\bibitem[{Parker} et~al., 1998]{parker98}
{Parker}, D.~G., {Ulrich}, R.~K., and {Pap}, J.~M. (1998).
\newblock {Modeling solar UV variations using Mount Wilson Observatory
  indices}.
\newblock {\em Solar Physics}, 177:229--241.

\bibitem[{Rabbette} and {Pilewskie}, 2001]{rabbette01}
{Rabbette}, M. and {Pilewskie}, P. (2001).
\newblock {Multivariate analysis of solar spectral irradiance measurements}.
\newblock {\em Journal of Geophysical Research}, 106:9685--9696.

\bibitem[{Schmahl} and {Kundu}, 1994]{schmahl94}
{Schmahl}, E.~J. and {Kundu}, M.~R. (1994).
\newblock {Solar cycle variation of the microwave spectrum and total
  irradiance}.
\newblock {\em Solar Physics}, 152:167--173.

\bibitem[{Schneider}, 2001]{schneider01}
{Schneider}, T. (2001).
\newblock {Analysis of Incomplete Climate Data: Estimation of Mean Values and
  Covariance Matrices and Imputation of Missing Values.}
\newblock {\em Journal of Climate}, 14:853--871.

\bibitem[{Schneider}, 2007]{schneider07}
{Schneider}, T. (2007).
\newblock {Comment on ''Spatio-temporal filling of missing points in
  geophysical data sets'' by D. Kondrashov and M. Ghil, Nonlin. Processes
  Geophys., 13, 151-159, 2006}.
\newblock {\em Nonlinear Processes in Geophysics}, 14:1--2.

\bibitem[{Tapping} and {Detracey}, 1990]{tapping90}
{Tapping}, K.~F. and {Detracey}, B. (1990).
\newblock {The origin of the 10.7 CM flux}.
\newblock {\em Solar Physics}, 127:321--332.

\bibitem[{Viereck} et~al., 2001]{viereck01}
{Viereck}, R., {Puga}, L., {McMullin}, D., {Judge}, D., {Weber}, M., and
  {Tobiska}, W.~K. (2001).
\newblock {The Mg II index: A proxy for solar EUV}.
\newblock {\em Geoph. Res. Lett.}, 28:1343--1346.

\bibitem[Wiener, 1964]{wiener64}
Wiener, N. (1964).
\newblock {\em {Extrapolation, Interpolation, and Smoothing of Stationary Time
  Series}}.
\newblock The MIT Press, Cambridge, Massachussets.

\bibitem[{Woods} et~al., 2000]{woods00}
{Woods}, T.~N., {Tobiska}, W.~K., {Rottman}, G.~J., and {Worden}, J.~R. (2000).
\newblock {Improved solar Lyman {$\alpha$} irradiance modeling from 1947
  through 1999 based on UARS observations}.
\newblock {\em Journal of Geophysical Research}, 105:27195--27216.

\end{thebibliography}

\end{document}